\documentclass[pdflatex,sn-nature]{sn-jnl}

\usepackage{graphicx}%
\usepackage{multirow}%
\usepackage{amsmath,amssymb,amsfonts}%
\usepackage{amsthm}%
\usepackage{mathrsfs}%
\usepackage[title]{appendix}%
\usepackage{xcolor}%
\usepackage{textcomp}%
\usepackage{manyfoot}%
\usepackage{booktabs}%
\usepackage{algorithm}%
\usepackage{algorithmicx}%
\usepackage{algpseudocode}%
\usepackage{listings}%
\usepackage{comment}%

\newcommand{\apj}{Astrophys. J.}   
\newcommand{\apjl}{Astrophys. J. Lett.}   
\newcommand{\apjs}{Astrophys. J. Suppl. Ser.}   
\newcommand{\aap}{Astron. Astrophys.}   
\newcommand{\icarus}{Icarus}   
\newcommand{\mnras}{Mon. Not. R. Astron. Soc.}   
\newcommand{\nat}{Nature} 
\newcommand{\pasp}{Publ. Astron. Soc. Pac.}   

\newcommand{\gretwo}{{\rm I\hspace{-1.2pt}I}}
\newcommand{\greone}{{\rm I}}

\newcommand{\farcs}{\mbox{\ensuremath{.\!\!^{\prime\prime}}}}

\begin{document}

\title[Article Title]{Winding Motion of Spirals in a Gravitationally Unstable Protoplanetary Disk }


\author*[1,2]{\fnm{Tomohiro} C. \sur{Yoshida}}\email{tomohiroyoshida.astro@gmail.com}

\author[1,2]{\fnm{Hideko} \sur{Nomura}}\email{hideko.nomura@nao.ac.jp}

\author[3]{\fnm{Kiyoaki} \sur{Doi}}\email{doi.kiyoaki.astro@gmail.com}

\author[4]{\fnm{Marcelo} \sur{Barraza-Alfaro}}\email{mbarraza@mit.edu}

\author[4]{\fnm{Richard} \sur{Teague}}\email{rteague@mit.edu}

\author[5,6]{\fnm{Kenji} \sur{Furuya}}\email{kenji.furuya@riken.jp}

\author[5,6]{\fnm{Yoshihide} \sur{Yamato}}\email{yyamato.as@gmail.com}

\author[7]{\fnm{Takashi} \sur{Tsukagoshi}}\email{takashi.tsukagoshi.astro@gmail.com}

\affil*[1]{\orgname{National Astronomical Observatory of Japan}, \orgaddress{\street{2-21-1 Osawa}, \city{Mitaka, Tokyo}, \postcode{181-8588}, \country{Japan}}}

\affil[2]{\orgdiv{Department of Astronomical Science}, \orgname{The Graduate University for Advanced Studies, SOKENDAI}, \orgaddress{\street{2-21-1 Osawa}, \city{Mitaka, Tokyo}, \postcode{181-8588}, \country{Japan}}}

\affil[3]{\orgdiv{Max Planck Institute for Astronomy}, \orgaddress{\street{Königstuhl 17}, \postcode{D-69117}, \city{Heidelberg}, \country{Germany}}}

\affil[4]{\orgdiv{Department of Earth, Atmospheric, and Planetary Sciences}, 
\orgname{Massachusetts Institute of Technology},
\orgaddress{\postcode{02139}, \city{Cambridge, MA}, \country{USA}}}

\affil[5]{\orgdiv{Department of Astronomy, Graduate School of Science}, \orgname{The University of Tokyo}, \orgaddress{ \city{Tokyo}, \postcode{113-0033}, \country{Japan}}}

\affil[6]{\orgname{RIKEN Pioneering Research Institute}, \orgaddress{\street{2-1 Hirosawa}, \city{Wako, Saitama}, \postcode{351-0198}, \country{Japan}}}

\affil[7]{\orgdiv{Faculty of Engineering}, \orgname{Ashikaga University}, \orgaddress{\street{268-1 Ohmae-cho}, \city{Ashikaga, Tochigi}, \postcode{326-8558}, \country{Japan}}}


\abstract{
The discovery of wide-orbit giant exoplanets has posed a challenge to our conventional understanding of planet formation by coagulation of dust grains and planetesimals, and subsequent accretion of protoplanetary disk gas.
As an alternative mechanism, the direct in-situ formation of planets or planetary cores by gravitational instability (GI) in protoplanetary disks has been proposed.
However, observational evidence for GI in regions where wide-orbit planets are formed is still lacking.
Theoretical studies predict that GI induces spiral arms moving at the local Keplerian speed in a disk.
Here, with multiple high angular resolution observations over a seven-year time baseline using the Atacama Large Millimeter/submillimeter Array (ALMA), we report the evidence for spiral arms following the Keplerian rotation in the dust continuum disk around the young star IM Lup.
This demonstrates that GI can operate in wide-orbit planet-formation regions, establishing it as a plausible formation mechanism for such planets.
}

\keywords{Protoplanetary disk, Planet formation}

\maketitle

\section{Introduction}\label{sec1}
Observations of exoplanets have revealed the existence of giant planets on orbits much wider than the Solar system planets \citep{akes13}.
For instance, in the HR 8799 system, four giant planets are detected at radii of 20-70 au from the central star \citep{maro08}.
The presence of the wide-orbit giant planets poses a challenge to planet formation theories.
According to the conventional bottom-up core accretion scenario, where the small dust grains grow with coagulation to create a planetary core and then accrete the gas in a disk \citep{poll96}, the formation timescale for giant planetary cores at these wide orbits is longer than the typical disk lifetime \citep{armitage2020}.
As an alternative mechanism, direct in-situ formation via disk fragmentation due to gravitational instability (GI) has been proposed \citep{boss97} although planet-planet/star scattering may be another possibility \citep{rode17}.
More recent simulations have suggested that full disk fragmentation is not necessarily required. Only marginally gravitationally unstable disks may efficiently form planetary cores, which could solve the time scale problem for the bottom-up core-accretion scenario \citep{rice04, bole10, baeh22, baeh23, long23a, long23b, rowt24}.
Therefore, it is critical to observationally reveal if GI can indeed happen in protoplanetary disks.

One of the most evident signatures of GI is spiral arms \citep[e.g.,][]{toom64, dong15}.
Indeed, such spiral arms have been observed in some protoplanetary disks \citep{muto12, huan18}.
However, spiral arms do not necessarily mean GI. Such spirals can also be induced by planet-disk interaction \citep{dong18}, stellar flybys \citep{cuel19}, and shadows cast by an inner disk \citep{mont16}.
Since their morphology resembles each other, it is generally challenging to distinguish those possibilities.

In particular, interpretations with GI or planet-disk interaction may result in different scenarios of planet formation.
Planet-induced spirals indicate an already-formed planet at a large radius, requiring planet formation to occur at an early evolutionary stage of the protoplanetary disk.
On the other hand, GI-induced spirals may suggest that the disk has enough gas for planet formation even at a relatively later stage, and planets can be formed there via GI.
A robust way to distinguish the two mechanisms is to characterize the motion of spirals.
Theories predict that the GI-induced spirals follow the local Keplerian motion \citep[e.g.,][]{coss09} (See also Method \ref{sec:theory} ), while the planet- (or flyby-) induced spirals follow the Keplerian motion at the companion's radius \citep{dong16}.
Recently, thanks to the increasing number of high-quality observations, it is possible to observe the motion of the spirals.
Indeed, efforts with high-resolution infrared imaging observations have been made \citep{ren20, xie23, ren24}. However, they suggested that the major spirals in the studied disks are induced by a fairly massive planet (or low-mass companion).
Another way to distinguish the scenarios is to investigate kinematical structures in spectral lines such as CO \citep{hall20, spee24}.
However, they trace the upper layer of the outer region in disks, which is different from the region where the wide-orbit exoplanets are detected.

In this article, we target a protoplanetary disk around IM Lup.
IM Lup is classified as a classical T Tauri star with a mass of $M_\star {\sim} 1.1\ M_{\odot}$ \citep{teag21} and located at 158 pc \citep{gaia2022} from the Earth.
The disk has been thoroughly studied, with numerous substructures characterized and stringent constraints on its physical properties being derived.
In particular, the mm-continuum emission was resolved at a very high angular resolution (${\sim}7$ au) by the DSHARP ALMA Large program \citep{andr18, huan18}, revealing grand-design spiral arms, and triggering follow-up studies to understand their origin.
In the infrared image, there is no significant shadow \citep{aven18}, excluding the shadow scenario.
Analysis of the rotation curves retrieved from CO line observations conducted by the MAPS ALMA Large Program \citep{ober21} obtained a kinematically-constrained disk mass of 0.1 $M_{\odot}$ \citep{loda23}.
Such a large disk mass can be converted to Toomre's Q \citep{toom64} ${\sim}1$, suggesting that the disk may be gravitationally unstable.
Additionally, there are other supports of GI in the IM Lup disk.
\cite{paneq24} found evidence for strong turbulence which can be induced by GI, although they used the CN lines tracing the elevated layer.
A vertical extension of the dust disk is also suggested which may also suggest strong turbulence \citep{bosm23}.
Further, the assumption that GI controls the gas surface density profile is consistent with the polarimetric infrared and mm-continuum images \citep{ueda24}.
On the other hand, other studies report kinematical detections of embedded Jupiter-mass planets, making alternative explanations for the spirals \citep{pint20, verr22}.
By distinguishing the two scenarios, GI or planet-driven, we can benchmark our current understanding of planet formation; can wide-orbit giant planets be formed via GI in a relatively-evolved Class II disk? Otherwise, is there an already-formed giant planet?
Furthermore, confirming one of the scenarios itself is important for characterizing the disk structure and physical processes \citep[e.g.,][]{ueda24}.

Thanks to the high interest, this disk has been observed several times with relatively high spatial resolutions using ALMA, spanning a period large enough to investigate the motion of the spirals.

\section{Detection of spiral motion}\label{main}
\subsection{Observations}\label{obs}

\begin{figure}[ht]
\centering
\includegraphics[width=\textwidth]{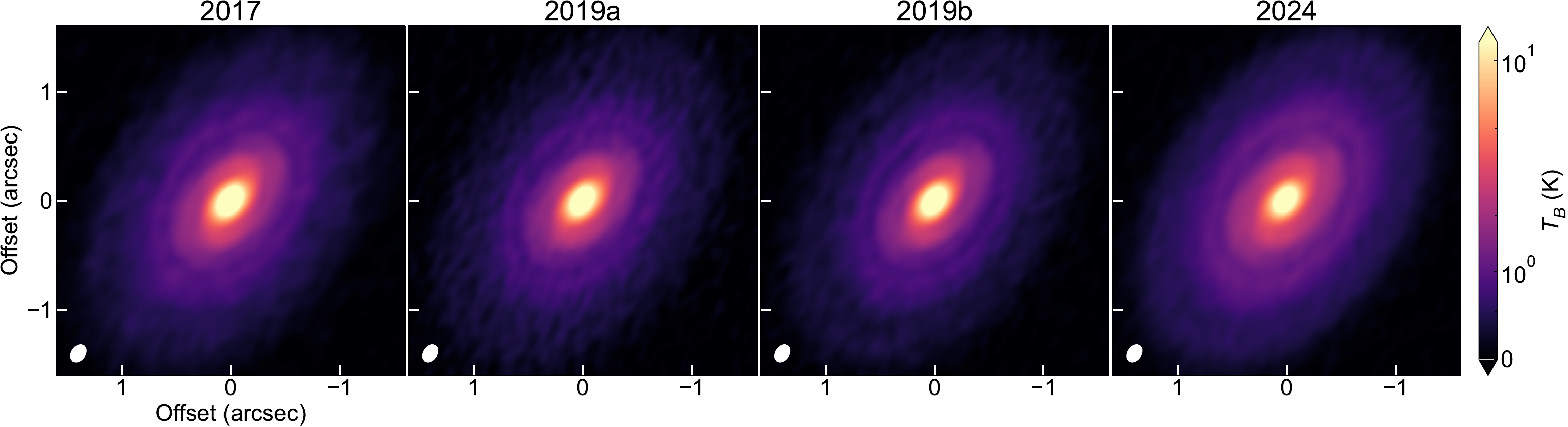}
\caption{Continuum images of the IM Lup disk in the three epochs used in the analysis. The color contour is shown in the brightness temperature. The white ellipses on the bottom left indicate the synthetic beam.}\label{fig_all}
\end{figure}

The IM Lup disk has been observed with a setup optimized for high angular resolution continuum observations in 2017, 2019, and 2024, resulting in a seven-year time baseline. 
These serve as the data used for this project.
For the years of 2017 and 2019, we use the archival ALMA data delivered from the DSHARP \citep{andr18} and MAPS \citep{ober21} collaboration, respectively.
The MAPS project observed two different frequency settings at 226 and 257 GHz in 2019.
Although their time baseline is negligibly small, we treat the two images in 2019 separately as epochs of 2019a (226 GHz) and 2019b (257 GHz).
The epoch of 2024, currently unpublished, was observed between Dec.~2023 and Aug.~2024 (PI: T. Yoshida, ID: \#2023.1.00525.S).
We describe the details of observations and data reduction in Methods.
We determine the representative date of each epoch as the mean observation date of the long-baseline observations because they have higher positional accuracy than the more compact configurations.
The key information of observations is summarized in Table \ref{tab1}.
\begin{table}[ht]
\caption{Observation Details}\label{tab1}%
\begin{tabular}{ccccccc}
\toprule
Epoch & Representative Date [MJD]  & Frequency & ALMA Band & Baselines & Project ID & Noise level \\
& & (GHz) & & (m) & & ($\rm \mu Jy\ beam^{-1}$) \\
\midrule
2017 & 2017 Oct 10\ [58036] & 238.984 & Band 6 & 15-13894 & 2016.1.00484.L \citep{andr18}  & 21\\
2019a & 2019 Aug 13\ [58708] & 226.053 & Band 6 & 15-3638 & 2018.1.01055.L \citep{ober21} & 28 \\
2019b & 2019 Aug 20\ [58715] & 257.029& Band 6 & 15-3396 & 2018.1.01055.L  \citep{ober21} & 26\\
2024 & 2024 Mar 22\ [60391] & 330.172 & Band 7 & 15-2517 & 2023.1.00525.S & 31 \\
\botrule
\end{tabular}
\footnotetext{The epoch of 2017 contains archival data from 2013.1.00226.S, 2013.1.00694.S, and 2013.1.00798.S \citep{ober15, clee17, pint18}, which were observed in 2014-2015. They are short baseline observations with short integration time, so it is unlikely that they trace the spiral structures. }
\end{table}
By adjusting the visibility weighting and convoluting a Gaussian, the beam sizes of all images are arranged to $150 \times 101$ mas with a position angle of 144.5 deg so that a circular beam with an FWHM of $150$ mas is obtained when the images are deprojected to face-on.
All the images used in the analysis are shown in Figure \ref{fig_all}.
In each epoch, spiral arms are robustly detected. 

\subsection{Dust rotation curve}\label{ana}
\begin{figure}[ht]
\centering
\includegraphics[width=\textwidth]{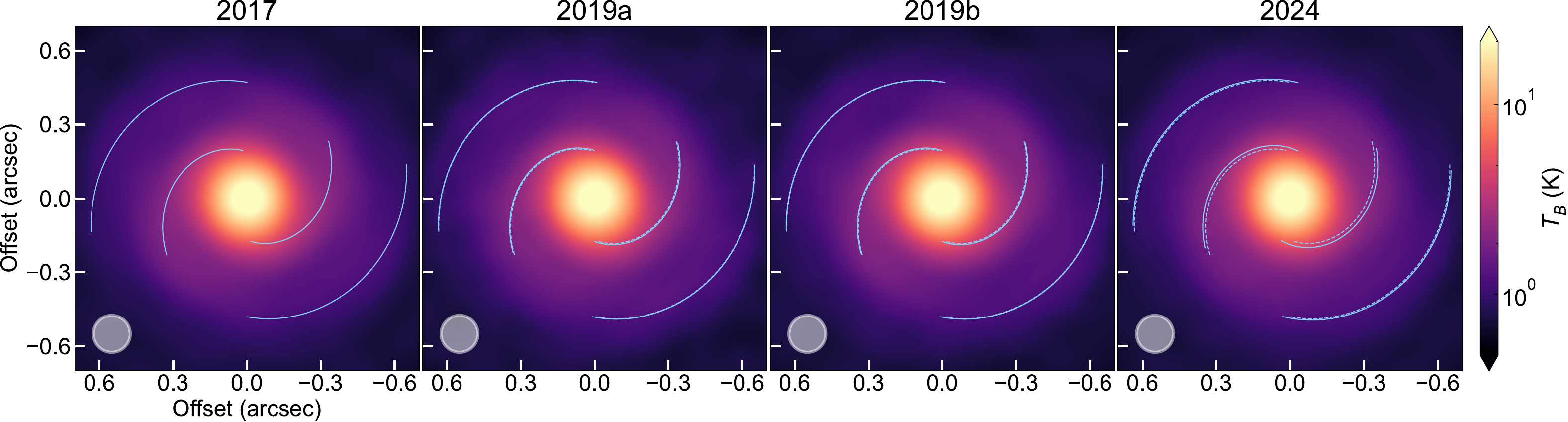}
\caption{Deprojected continuum images used in the quantitative analysis.
These images are normalized based on the 2017 epoch and shown in the unit of the corresponding brightness temperature (see Methods).
The blue solid curves indicate the spiral ridge \citep{huan18} at each epoch, assuming that they follow the local Keplerian rotation (i.e. the case of GI).
The blue dashed curves correspond to the spiral ridge in 2017.
The circles on the bottom left indicate the synthetic beam after deprojection.
}\label{deproj}
\end{figure}
\begin{figure}[ht]
\centering
\includegraphics[width=\textwidth]{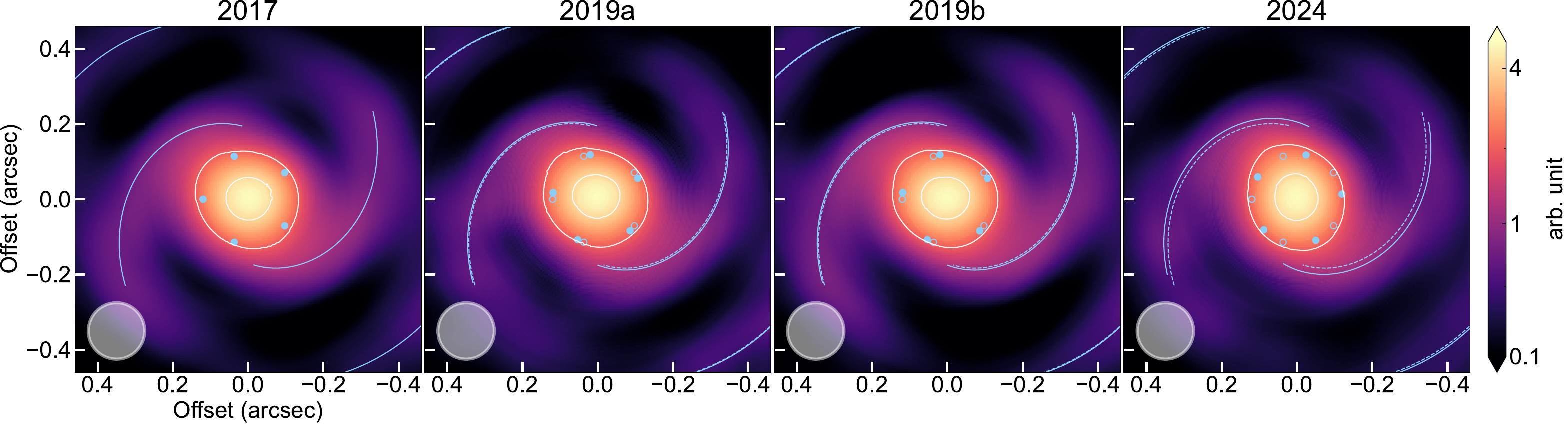}
\caption{Deprojected continuum images after partially subtracting the axisymmetric component.
The color scale is also stretched for visibility purposes (see Extended data for details).
The original data is the same as Figure \ref{deproj}.
The white contours are drawn at values of 2 and 4 (in an arbitrary unit).
The blue-filled scattered points follow the local Keplerian rotation, while the blue circles show the positions in 2017 for comparison. The rotation of the contour is consistent with the local Keplerian rotation.
}\label{deproj2}
\end{figure}
We deproject the images with known geometrical parameters \citep{ober21, teag21}, correct the intensity difference among frequencies as explained in Methods, and plot in Figure \ref{deproj}.
In Figure \ref{deproj2}, the same data as Figure \ref{deproj} is plotted, but the spirals are emphasized by partially subtracting the axisymmetric component (see Extended Data).
Here, we plot the spiral ridges specified by \cite{huan18}, which uses the original high-resolution DSHARP image taken in 2017, with the temporal variations of the ridges expected for the local Keplerian rotation.
In addition, we add points following the local Keplerian rotation at $r=0\farcs12$ as a guide.
It is clear that there is a temporal variation in the observed images, and they are visually consistent with the local Keplerian rotation, especially in the inner region.
This can also be seen in an animation based on Figure \ref{deproj2} (Extended Data).

To quantitatively analyze the time variability of the three images, we introduce two approaches as described in detail in Methods.
In the first approach (``Image-fitting analysis'', see Method \ref{image_fitting}), we assume a parametrized angular velocity profile, $\omega(r)$, where $\omega(r)$ is a power-law of a radius $r$ with a constant floor.
The original high-resolution DSHARP image \citep{andr18} is azimuthally shifted according to the angular velocity profile and the time separation, and compared with the last three epochs after beam-convolution.
On the other hand, in the second approach (``Radius-by-radius analysis'', see Method \ref{radius_radius}), we azimuthally shift the images radius-by-radius and find the optimal azimuthal shift between images.
Although this approach may be affected by the limited spatial resolution, it is advantageous that no assumption of the angular velocity profile is needed.

\begin{figure}[ht]
\centering
\includegraphics[width=\textwidth]{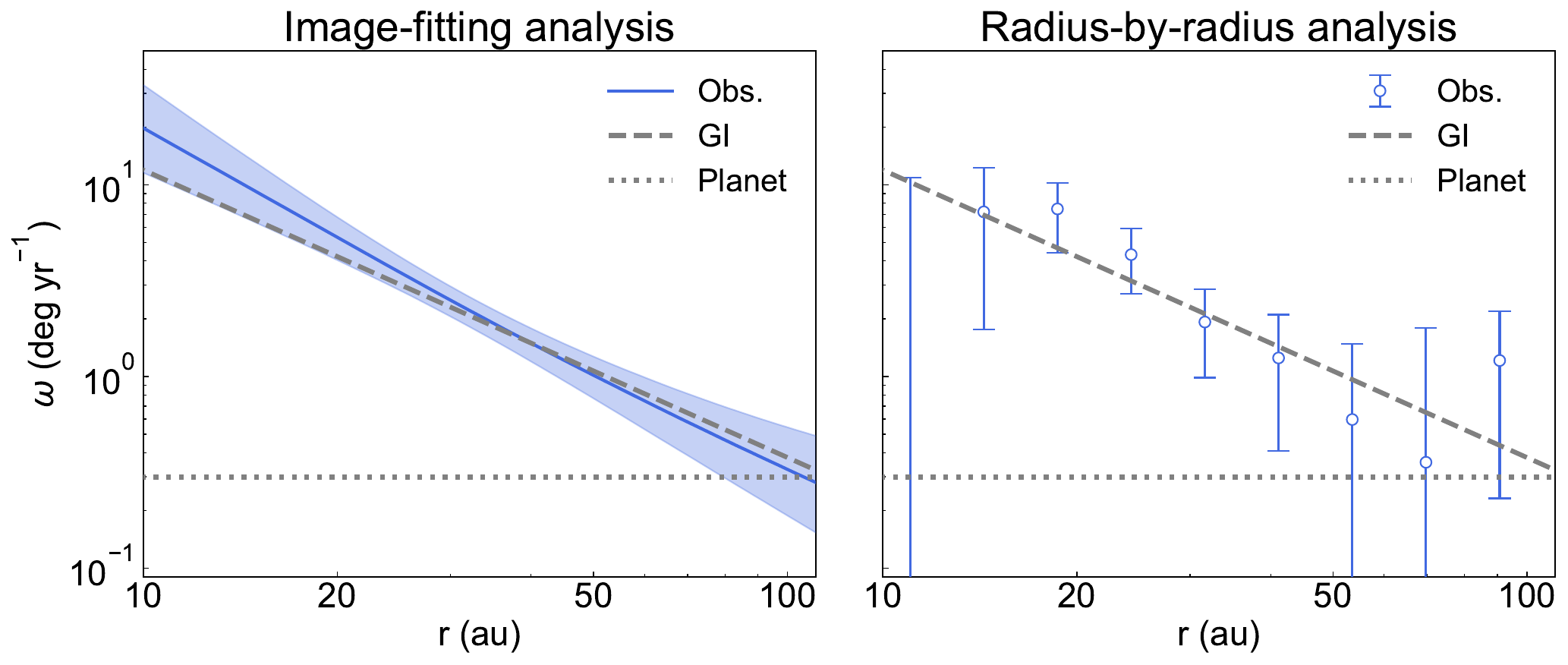}
\caption{Rotation curve of the dust disk. $\omega(r)$ extracted with the image-fitting analysis is plotted in the left panel (blue line) while the radius-by-radius analysis result is shown in the right panel (blue circles). In the left panel, the blue solid line and the blue-shaded regions indicate the median and 16-84th percentiles of the marginal probability distribution at each radius, respectively.
The error bars in the right panel correspond to the $1\sigma$ range of the normalized probability distribution at each radius. Note that each point is not perfectly independent as the beam size is ${\sim}24$ au.
The dashed and dotted lines, respectively, show predictions for spirals induced by GI and a planet located at 117 au from the central star \citep{pint20}. 
}\label{rot_curve}
\end{figure}

The resultant dust rotation curve is plotted in Figure \ref{rot_curve}.
The motion of the spirals is successfully detected in both approaches, where they agree well with each other.
The dust rotation curve exactly matches the local Keplerian rotation, while it is significantly different from the rigid body rotation.
As described in the introduction, the detection of the local Keplerian motion of the spirals directly suggests that these spirals are being driven by GI.

\section{Implications and Future Direction}\label{implications}

The spirals in the dust continuum image are evident from $r<20$ au up to $r {\sim} 200$ au \citep{huan18}.
This study suggests that GI can happen in the primary planet-forming region, where most planets are thought to be formed.
GI happens if the gas surface density is sufficiently large and Toomre Q is comparable or smaller than unity \citep{toom64}.
Indeed, the disk mass measurement using the CO rotation curves implies $Q{\sim}1.0-2.5$ across the whole disk \citep{loda23}.

The winding motion suggests that these spirals are transient.
The pitch angle may soon become very small, diminishing the spirals in the Keplerian time scale.
However, there are several spirals up to $r {\sim} 200$ au with moderate pitch angles of ${\sim}10-20^\circ$ \citep{huan18}.
Unless all spirals are formed very recently at the same time, it is more reasonable to consider that the spirals are continuously formed and diminished.
This may indicate that the disk is gravitationally self-regulated, maintaining the marginally unstable state with $Q {\sim} \mathcal{O}(1)$.
This scenario is also supported by the successful reproduction of multi-band continuum observations under the assumption of gravitational self-regulation \citep{ueda24}.
Gravitational self-regulation implies that the disk does not cause fragmentation on a short time scale, which is consistent with the fact that two-armed left-right symmetric spirals are evident in the IM Lup disk.
This is in contrast to another disk with confirmed GI, AB Aur, where much more complex structures are observed \citep{spee24}.

Such marginally (un)stable conditions may accelerate the formation of planetary cores \citep{rice04, bole10, baeh22, baeh23, long23a, long23b, rowt24}.
If this happens in this radius range, it can be directly compared with detected wide-orbit giant planets as seen in the HD 8799 system \citep{maro08}.
On the other hand, the direct formation of giant planets by disk fragmentation in the future requires a shorter cooling time scale and/or a more massive disk.
This route may be possible if more material is supplied from the interstellar medium by late-stage infall as recently suggested in the Lupus star-forming region, where IM Lup resides \citep{wint24a, wint24b}.
Indeed, theoretical studies suggest that two-armed spirals can be more likely dominant when there is significant mass infall onto the disk \citep[e.g.,][]{hars11, tsuka11, tsuka15, tomi17}(see Methods \ref{sec:theory}), which is consistent with the IM Lup disk.
Another system with a confirmed GI, AB Aur\citep{spee24}, also exhibits infall streamers \citep{spee25}.
It is implied that late infall from the ISM may play a significant role in the evolution and dynamics of protoplanetary disks.
It is notable that such disk fragmentation may also produce not only giant planets but also brown dwarf companions \citep{krat10}.

Thanks to the growing number of archival data, time-domain astronomy is becoming available even in the field of planet formation \citep[e.g.,][]{teag22, debe23, kuo24}.
Regarding the IM Lup disk, although we already demonstrated that we can even model-independently determine the dust rotation curve for each radius, the accuracy becomes higher for a longer time baseline and higher spatial resolution.
For example, in the next ALMA cycle 12 (2026), it is expected that spirals at $r=40$ au will have moved by ${\sim}15^\circ$ or $70$ mas compared to the DSHARP image.
This can be perfectly resolved with the beam size of the original DSHARP image ($44$ mas).
As the signal-to-noise ratio of the spiral arm contrast exceeds ${\sim}10$ at this radius \citep{huan18}, the positional accuracy of ${\sim}5$ mas is achievable \citep{alma_tech}.
Therefore, if an image that has comparable quality to DSHARP is available, the spiral speed between the two epochs can be determined with an accuracy of ${\sim}10\%$.
This kind of high-accuracy measurement of the dust rotation curve will bring a unique opportunity to investigate disk structure in detail.
For example, we will be able to detect if the GI spiral propagation speed deviates from the Keplerian speed by more than 10 \%, further constraining the disk mass \citep{coss09}.

Overall, confirmation of GI makes the primary planet-forming region in the IM Lup disk an excellent laboratory for studying protoplanetary disks and planet formation.

\section{Methods}\label{sec11}

\subsection{Observation details, data reduction and imaging}

\subsubsection{2017 image}
For the first epoch, we used the self-calibrated DSHARP measurement set available on their website \citep{andr18} and performed imaging by employing the CLEAN algorithm with the $\tt mtmfs$ deconvolver in the {\tt tclean} task of the Common Astronomy Software Applications (CASA, version 6.5.5 \citep{CASA}).
Since this epoch has larger baselines compared to other epochs (Supplementary Table \ref{tab:image}), for a fair comparison, we employed ${\tt uvtaper} = 0\farcs08$. 
Several weighting schemes were tested to reduce the noise level with the required beam size being kept.
We found that the super-uniform weighting with ${\tt npixels=10}$ achieves this.
As a CLEAN mask, we adopted an elliptical mask with a semi-major axis of $2.''0$ with the geometrical parameters of the IM Lup disk (the position angle of 144.5 deg and inclination angle of 47.5 deg \citep{ober21, teag21}), which were also used in other epochs.
The resultant beam size was $81 \times 81$ mas with a position angle of $41$ deg and the RMS noise level was 18 ${\rm \mu Jy\ beam^{-1}}$.
The image was re-convolved with a Gaussian to obtain a common beam size of $150 \times 101$ mas. 

\subsubsection{2019 images}
The MAPS dataset was adopted for the epochs of 2019a and 2019b \citep{ober21, sier21}.
The original continuum visibility data were kindly provided by the MAPS collaboration.
We made an image with a robust parameter of $0$, resulting in a beam size of $110 \times 80$ mas with a position angle of $-81$ deg.
The RMS noise level was estimated to be $28$ and $27\ {\rm \mu Jy\ beam^{-1}}$ for 2019a and 2019b, respectively.
We reconvolved these images in the same way as the 2017 image.

\subsubsection{2024 image}
In ALMA Cycle 10 (\# 2023.1.00525.S, PI. T. Yoshida), the IM Lup disk was newly observed with two different spectral setups.
The first setup was centered ${\sim}339$ GHz and observed on Dec. 9, 2023, Dec 15, 2023, Jun. 27, 2024, and Jul. 05 2024 with an extended configuration (C43-6), and on Apr. 30, 2024, and Aug. 24 2024 with a compact configuration (C43-3). The total integration time was 3.8 hours.
The second setup was centered ${\sim}321$ GHz and observed on May. 29, 2024 with the configuration C43-4, totaling 1.4 hours of integration.

The two setups were separately self-calibrated after running the pipeline calibration.
Note that we did not apply the automated self-calibration but manually performed it.
For the first setup, the spectral windows were originally configured to observe the 331 GHz continuum emission, $\rm ^{12}CO\ J=3-2$, and $\rm ^{13}CO\ J=3-2$ lines.
We used the 331 GHz continuum and $\rm ^{12}CO\ J=3-2$ spectral windows in the following analysis because they dominate the effective total bandwidth.
Note that the $\rm ^{12}CO\ J=3-2$ line was flagged here and will be discussed in a future study.
We performed five rounds of iterative phase-only self-calibration for the short-baseline data.
Solution intervals were set to the duration of execution blocks in the first round, and shortened to $80$ s in the final round.
Then, we combined the self-calibrated short-baseline with the long-baseline data.
Additional eight rounds of phase-only self-calibration with solution intervals of the duration of the execution blocks to 20s, and one round of amplitude self-calibration with a solution interval of the execution block duration were applied.
The RMS noise level was improved by a factor of 1.7 before and after self-calibration of the combined data set.
The second setup targeted the 315 and 328 GHz continuum emission, $\rm C^{18}O\ J=3-2$, and $\rm ^{13}C^{18}O\ J=3-2$ lines.
We used all spectral windows after flagging the lines.
We performed seven rounds of phase-only self-calibration with the solution intervals from the execution block (EB) length to 15 s and one round of amplitude self-calibration with the solution intervals of the EB length.

After combining all the data, we additionally ran one round of the phase-only self-calibration to align the phase center.
Finally, we CLEANed the data and obtained the image at a representative frequency of 330.172 GHz.
We adopted a robust parameter of 0 to achieve the required beam size without significantly losing the sensitivity, while that of 0.5 was used during self-calibration.
The resultant synthesized beam size was $110 \times 90$ mas with a position angle of $-77$ deg.
The RMS noise level was estimated to be $29\ {\rm \mu Jy\ beam^{-1}}$.
The image was reconvolved to have the common beam.
Imaging parameters are summarized in Supplementary Table \ref{tab:image}.

\subsection{Analysis}

\subsubsection{Image normalization} \label{imnorm}

The four images are observed at different frequencies.
The longest time baseline images (2017 and 2024) have a frequency separation of ${\sim}30\%$. 
Additionally, the absolute flux uncertainty of ALMA might alter the intensity.
Therefore, in advance of detailed analyses, we normalize the images to fairly compare the four images.

First, all images are positionally aligned by specifying the disk center by fitting an elliptical Gaussian to the central bright component.
Then, we deproject the images by adopting the measured geometrical parameters of the IM Lup disk, which were also used as the CLEAN mask (the position angle of 144.5 deg and inclination angle of 47.5 deg \citep{ober21, teag21}).
We azimuthally average the de-projected images and obtain the radial intensity profile.
At each radius, the observed images are divided by a ratio of their azimuthally averaged profile to that of the 2017 image.

We then iteratively determine the disk center in the following way.
First, we perform the image-fitting analysis (Sec \ref{image_fitting}), where we have parameters for a disk offset from the image center.
The obtained offsets are then used to deproject the images again.
The deprojected images are then repeatedly used for the image-fitting.
We iterate this procedure four times and check the convergence of the offset parameters.

This normalization enables us to focus on only the azimuthal variation.
The normalized images indeed resemble each other except for a small rotation of the pattern (Figures \ref{deproj} and \ref{deproj2}).
An animation made from these normalized images is presented in Extended Data.

\subsubsection{Image-fitting analysis} \label{image_fitting}
In this analysis, we first prepare the original high spatial resolution ($44\times43$ mas, PA$=115^\circ$) DSHARP image that is available as a fits file on the official website as a reference image.
The image is then convolved to have a circular beam with an FWHM of 65 mas after deprojection, which is more than twice as good as the common resolution, and deprojected using the geometrical parameters.
Here, the disk center is specified by 2D-Gaussian fitting to the central bright component.

We assume that the disk pattern motion follows
\begin{equation} \label{eq:omega}
    \omega(r) = \omega_0 \left( \frac{r}{1''} \right)^{-\gamma} + \omega_f,
\end{equation}
where $\omega_0$ and $\omega_f$ control the absolute value of angular velocity while $\gamma$ indicates the radial dependence of it.
These quantities are treated as free parameters.
This parametric form is an assumption, but can be fitted to both cases: GI- and/or planet-induced spirals.
The deprojected reference image is rotated according to Equation (\ref{eq:omega}) and convolved with a Gaussian again to match the common beam size of 150 mas.
Then, we compare them with the 2019a, 2019b, and 2024 images in terms of a log-likelihood function,
\begin{equation} \label{eq:loglike}
    -\frac{1}{2} \sum_{i} \sum_{\rm image} \left( \frac{ I_i - I_{\rm ref} }{\sigma_{i - {\rm ref}}} \right)^2,
\end{equation}
where $I_i$ is the images in the epoch $i$, $I_{\rm ref}$ is the reference image after rotating and parallel-shifting by $(dx, dy)$ for each epoch.
Although we spatially align the images in advance (Sec \ref{imnorm}), we can simultaneously treat the effect of the alignment accuracy by taking the offset $(dx, dy)$ as free parameters.
The summation $\sum_{\rm image}$ is taken on the spatial domain.
We only sum up the radial range of $r=0\farcs065-0\farcs7$, where the inner edge is limited by the beam size of the DSHARP image and the outer edge is limited by the spiral detection \citep{huan18}.
$\sigma_{i{\rm-ref}}$ is the standard deviation of a residual map between the epoch $i$ and the reference image.
To sample the posterior distribution of the free parameters, we use a Python package {\tt emcee} \citep{emcee} in practice.
We adopt broad non-informative priors (Supplementary Table \ref{tab2}).
We employ 128 walkers and 5000 steps, with the first 3000 steps discarded as burn-in.

The resulting percentiles and best-fit values that maximize the posterior probability are tabulated in Supplementary Table \ref{tab2}.
The rotation curve with uncertainties is estimated to be 
\begin{equation}
\omega(r) = 0.11^{+0.12}_{-0.06} \left( \frac{r}{1''} \right)^{-1.9^{+0.44}_{-0.49}}\ {\rm deg\ yr^{-1}}.
\end{equation}
The floor term ($\omega_f$) is not well constrained but only an upper limit is obtained, which is negligible at the radial region of interest (see Figure \ref{rot_curve}).
Using the stellar mass of $1.1\ M_\odot$ and considering the stellar gravity, the Keplerian velocity profile is
\begin{equation} \label{omega_kep}
\omega_{\rm kep}(r) \simeq 0.19 \left( \frac{r}{1''} \right)^{-1.5}\ {\rm deg\ yr^{-1}},
\end{equation}
which is consistent with the derived profile within the uncertainty.
Note that we neglect the self-gravity of the disk.
On the other hand, the rigid body rotation expected for the planet-induced spiral is
\begin{equation} \label{omega_planet}
\omega_{\rm planet}(r) \simeq 0.30 \ {\rm deg\ yr^{-1}},
\end{equation}
which differs from the constrained rotation curve by at least $4\sigma$ in $\gamma$.

In Supplementary Figure \ref{fig:corner}, we plot the marginal probability distributions.
The case for GI is shown with the blue lines while the planet-induced case is marked with the orange lines \citep{pint20}.
We perform the kernel density estimate on the MCMC sample with {\tt scipy.stats.gaussian\_kde} \citep{scipy} and calculate the probability normalized by the maximum value.
The parameters of the local Keplerian rotation (Eq. \ref{omega_kep}) and the Keplerian rotation at 117 au (Eq. \ref{omega_planet}) return $0.7$ and $7\times10^{-50}$, respectively.
These correspond to 0.9$\sigma$ and 15$\sigma$, respectively, showing that the GI-induced scenario is significantly preferable compared to the planet-induced scenario.

\subsubsection{Radius-by-radius analysis} \label{radius_radius}
In the image-fitting analysis (Sec \ref{image_fitting}), we assumed that the radial rotation profile is described as Equation \ref{eq:omega}.
To check if there is a non-power-law motion, we performed an analysis for each radius in this subsection.
As a function of radius $r$ and azimuthal angle of $\theta$, the intensity profile of the normalized epoch $i$ image can be expressed as $I_i(r, \theta)$.
We consider all combinations in the four images and calculate a residual between a pair.
In practice, we define a function
\begin{eqnarray}
\chi^2(r, \omega) = \sum_{i<j} \sum_\theta \left( \frac{ I_i(r, \theta) - I_j(r, \theta - \omega \Delta T_{i,j} ) }{\sigma_{i,j}} \right)^2
\end{eqnarray}
where $\Delta T_{i,j}$ is a time separation between epoch $i$ and $j$, and $\omega$ is the angular velocity that we aim to constrain at each $r$.
$\sigma_{i,j}$ is given as
\begin{equation}
\sigma_{i,j} = \sqrt{ \sigma_i^2 + \sigma_j^2 },
\end{equation}
where $\sigma_i$ is the standard deviation of epoch $i$.
At each $r$, we calculate $\chi^2$ for various $\omega$ and find an optimal $\omega$ that minimizes $\chi^2$.
The optimal $\omega$ indicates the averaged angular velocity among three epochs.
This method is straightforward, however, it is difficult to accurately estimate the uncertainty of $\omega$ if there is a signal correlation with intensity gradient across radii.
Indeed, the spatial separation of the spiral in one epoch is similar to the beam size in our images, implying this is the case.
However, it would be useful to calculate the $1\sigma$ uncertainty from the parameters where $|\chi^2 - {\rm min ( \chi^2 )|} = 1$, which is provided as error bars in Figure \ref{rot_curve}.
Here, the radial bin is taken logarithmically.
In addition, we present the map of the normalized probability distribution for each pair and plot in Supplementary Figure \ref{fig:epcomp}.

\subsubsection{The quantity we measured}\label{sec:measured}
In the methods presented in this paper, we measure the pattern speed using all azimuthal information at each radius.
This means that we do not assume any radial structure such as disk-global spiral arms.
Therefore, even if there is temporal variation in the radial structure such as disconnection of spirals (see Section \ref{sec:theory}), the pattern speed can be measured.

Still, we have to care about what we exactly trace.
Since the observations have a limited spatial resolution, we might be biased to the spirals with low azimuthal modes ($m$) due to smoothing out \citep[e.g.,][]{dipi14}.
However, in the case of the IM Lup disk, the high spatial resolution image \citep{huan18} showed that the $m=2$ is likely the dominant mode because the detected number of spiral arms is only two even in the well-spatially resolved region ($r \gtrsim 50\ {\rm au}$).
Therefore, what we measure in our analysis is the pattern speed of the dominant $m=2$ spiral arms.

\subsection{Comments on spiral pattern speed expected for GI} \label{sec:theory}
The speed of GI-induced spiral patterns has been investigated via hydrodynamic simulations.
A three-dimensional simulation by \cite{coss09} showed that the spiral pattern speed is roughly Keplerian at each radius via Fourier analysis of all azimuthal modes. The small deviation from the Keplerian speed ($<15\%$) depends on the disk-to-stellar mass ratio $q$.
\cite{beth21} performed a more detailed analysis of the temporal variability of the spiral structure in self-gravitating disks.
Interestingly, in a run with $q=1/3$, they found that the spiral wakes radially disjoint and meet another wake as they rotate.
This means that large-scale spirals only appear momentarily or stochastically (see their Fig. 10).
In this case, we cannot define the pattern speed of individual spiral arms; observable large-scale spiral structures may continuously appear and disappear.
However, their Fourier analysis suggested that the pattern speed computed from all azimuthal modes is roughly Keplerian as \cite{coss09} found.
They also performed Fourier analysis by only focusing on the dominant modes of $2\leq m \leq6$, showing that the rotation profile is ``piecewise'' rigid body rotation. However, the deviation from the Keplerian speed is still moderate and up to $30\%$ (see their Fig.11).
Additionally, they ran simulations with various $q$.
In the case of $q=1$, they found that the dominant azimuthal mode is $m=2$ and their pattern speed is close to the Keplerian speed at every radius.
It is notable that the dominant azimuthal mode is roughly scaled as $m{\sim}1/q$ \citep{dong15}, implying that $q{\sim}0.5$ might be enough to see a similar situation.
We also note that these results are somewhat different from the classical linear theory with many simplifications \citep[e.g.,][]{linshu64}, where each spiral is specified with the mode numbers and rigidly rotates at a specific angular speed.

These simulations assume that the disk is well-isolated from the interstellar medium (ISM).
However, it is known that material infall onto the disks may affect the GI-induced spirals.
For example, \cite{hars11} found that infall makes low-order, global modes of spiral arms.
Even in the case with a moderate $q$ of 0.1, low-order modes that are otherwise only seen for $q=0.2$ are observed.
\cite{tsuka15} also showed that $m=2$ is the dominant mode of a disk with infall.
\cite{tomi17} performed three-dimensional magneto-hydrodynamic simulations of a disk in a collapsing molecular cloud core and presented that grand design $m=2$ spiral arms are generated.
Here, the mass accretion rate onto the disk from the envelope is one order of magnitude higher than that onto the star from the disk.
They found that the spiral arms rotate at the local Keplerian speed.
The spiral arms wind up on a Keplerian time scale but recurrently form because of the mass loading due to infall, resulting in a high spiral occurrence probability of ${\sim}50\%$.
This picture can be clearly seen in an animation they provided.

In summary, these theoretical studies suggest that $m=2$ spiral arms moving at the local Keplerian speed should be observed when $m=2$ is the dominant mode.
This is possible when the disk is massive and/or there is a significant infall onto the disk from the ISM.
Note that, if a larger $m$ is dominant rather than $m=2$, the pattern speed of the $m=2$ spiral may deviate from the local Keplerian since the average speed of all modes is almost Keplerian \citep{coss09, beth21}.

In the IM Lup disk, $m=2$ is likely the dominant mode (Sec \ref{sec:measured}).
If we consider an isolated disk, this is inconsistent with the disk mass of $q\simeq0.1$ suggested by the CO rotation curve analysis \citep{loda23}.
There may be two ways to solve the paradox.
First, the disk mass or surface density at the region we analyzed could be larger than a disk with $q\simeq0.1$, given that there might be some level of uncertainty on the kinematic measurement \citep{andr24}.
Second, there may be significant infall onto the IM Lup disk from the ISM, making low modes more dominant \citep{hars11}.

The infall scenario is in line with the recently proposed paradigm, where significant mass infall due to the Bondi-Hoyle-Lyttleton (BHL) accretion is needed to explain disk characteristics even in the Class II stage \citep{pado24, wint24a}.
Indeed, \cite{wint24b} found that the stellar accretion rate is spatially correlated in the Lupus star-forming region where IM Lup resides, suggesting that mass accretion from the ISM is effective.
Based on their estimate, the BHL accretion rate onto the disks can be ${\sim}10$ times higher than the stellar accretion rate, which may be a preferable condition for the GI induced by infall \citep{hars11, tomi17}.
In addition, it is theoretically suggested that GI-induced spiral arms are observable only in the first $10^4$ yrs in an isolated system \citep{hall19}.
Given the age of the Lupus star-forming region (${\sim}1-3\ {\rm Myr}$\citep{galli20}), it requires significant mass loading in the late stage.

Although there has been no direct evidence of infall onto the IM Lup disk itself, there is a possible infall motion detected in the CO line in the RU Lup disk which is located at just ${\sim}10$ arcmin or ${\sim}0.5$ pc from IM Lup on the plane of the sky \citep{huan20}.
In another system with a confirmed GI, AB Aur \citep{spee24}, \cite{spee25} detected infalling gas and concluded that the infall induces GI.
Future sensitive observations of gas around IM Lup are the key to testing this picture.

\backmatter

\section{Extended Data}
We provide the animation of the spiral motion (Supplementary Video).
For visibility purposes, the spiral structure is emphasized in the following way.
First, we extracted the radial intensity profile by taking the azimuthal average of the normalized images.
Then, original images with $20\%$ enhanced intensities are subtracted by axisymmetric images created using the radial profile so that we can easily see the non-axisymmetric structures together with the radial dependence of the intensity.
We also plot the spiral ridges specified by \cite{huan18} with the temporal variations of the ridges expected for the local Keplerian rotation as in Figure \ref{deproj}.

\section*{Declarations}

\begin{itemize}

\item Data availability:
The data used for this study are available on the ALMA science archive (\url{https://almascience.nao.ac.jp/aq/}).
The project IDs are 2016.1.00484.L, 2013.1.00226.S, 2013.1.00694.S, 2013.1.00798.S, 2018.1.01055.L, and 2023.1.00525.S.

\item Code availability:
The codes used for the analysis are available from the corresponding author upon reasonable request.

\item Acknowledgements:
We thank the anonymous referees for their helpful comments that improved the manuscript.
We also thank Dr. Kengo Tomida for the helpful discussion on the spiral speed.
We acknowledge Dr. Anibal Sierra for providing the calibrated MAPS continuum data.
This paper makes use of the following ALMA data: 2016.1.00484.L, 2013.1.00226.S, 2013.1.00694.S, 2013.1.00798.S, 2018.1.01055.L, and 2023.1.00525.S.
ALMA is a partnership of ESO (representing its member states), NSF (USA) and NINS (Japan), together with NRC (Canada), NSTC and ASIAA (Taiwan), and KASI (Republic of Korea), in cooperation with the Republic of Chile. The Joint ALMA Observatory is operated by ESO, AUI/NRAO and NAOJ.
This work was supported by Grant-in-Aid for JSPS Fellows, JP23KJ1008 (T.C.Y.) and JSPS KAKENHI grant Nos. 19K03910, 20H00182 (H.N.).

\item Author contribution:
T.C.Y. led the observing proposal, data processing, analysis, and manuscript preparation.
R.T. and M.B. were involved in the discussion of analysis and results as well as manuscript preparation.
H.N., K.D., K.F., Y.Y., and T.T. were participants in writing the observing proposal and manuscript.

\item Competing Interests:
The authors declare no competing interests.

\end{itemize}

\noindent



\newpage

\section{Supplementary Information}
\renewcommand{\figurename}{Supplementary Figure}
\renewcommand{\tablename}{Supplementary Table}
\setcounter{figure}{0}
\setcounter{table}{0}

\subsection{Frequency effect} \label{freq_effect}
The analyzed images have different representative frequencies.
Throughout the analyses, we assumed that the different frequencies trace the same dust surface density structure, and therefore, the apparent difference observed between different epochs is temporal variation.
If this assumption were not valid, the frequency difference could artificially produce the temporal variation.
\begin{figure}[ht]
    \centering
    \includegraphics[width=0.9\textwidth]{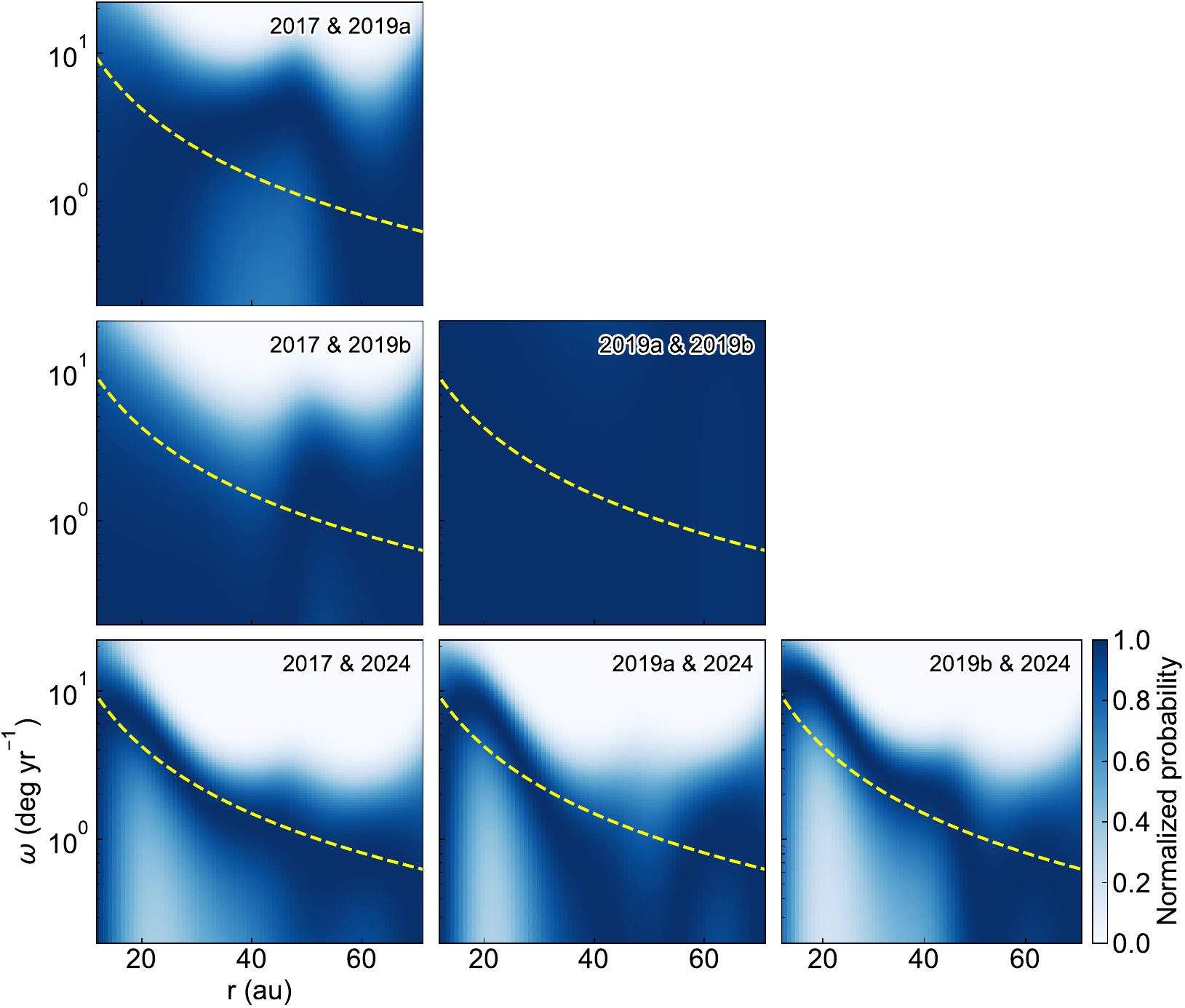}
    \caption{Normalized probability maps of $\omega$ for each pair of epochs.
We calculated $\exp({-\chi^2/2})$ at each radius for a pair of epochs, and normalized it by the maximum value at each radius. 
The yellow dashed lines indicate the local Keplerian angular frequency.}
    \label{fig:epcomp}
\end{figure}
To test this based on the data, we plot the normalized probability maps for each epoch in Supplementary Figure \ref{fig:epcomp}.
Although the uncertainties are different depending on the panels, the optimized $\omega$ curves are similar independently of the choice of epochs. 
In particular, for example, although the 2019b image has a frequency offset of $14\%$ compared to the 2019a image, the pairs of 2019a \& 2024 and 2019b \& 2024 are similar to 2017 \& 2024.
This means that the frequency effect is minimal.

In addition to the data-oriented approach, we theoretically investigate whether the assumption is valid or not in two different cases where the dust emission is optically thick or thin.

\subsubsection{ Optically thick case }
If the dust emission is optically thick and the disk has a non-zero inclination, the observed positional difference could be attributed to the different emitting surfaces between the two frequencies.
This effect can be seen in a comparison of a spiral in infrared and millimeter wavelength images of a protoplanetary disk, where the emitting layers are significantly different \citep{brow21}.
However, given that our images are both in millimeter wavelengths, which trace the relatively large dust grains near the midplane, the effect should be small, as shown in the following.
Assuming an extreme case where the longer wavelength image traces the midplane and the shorter wavelength one traces the upper layer at $z$ au from the midplane, the projected difference of the same surface density structure between the two wavelengths can be expressed as $\Delta X_{\rm emis} = z \tan i$,
where $i$ is the inclination angle of the disk.
$z$ can be approximated by the dust scale height, which is typically less than a few percent of the gas scale height \citep[e.g.,][]{vill20, vill22} and at most comparable to it \citep{doi21}.
If we assume the ultimate case where the dust scale height is similar to the gas scale height and set $z = 0.05r$ \citep{ueda24}, we can estimate an upper limit of $\Delta X_{\rm emis}$ as
\begin{equation}
    \Delta X_{\rm emis} = 0\farcs017 \left ( \frac{r}{50\ {\rm au}} \right ).
\end{equation}
This is sufficiently smaller than the observed positional difference at 30 au (${\sim}0\farcs05$), suggesting that the frequency effect is negligible.
Furthermore, we note that this effect should make the positional difference in the same direction with respect to the disk's major axis. This is different from what is expected for rotation and suggested observationally.

\subsubsection{ Optically thin case }
In the case where the dust emission is optically thin, the observed positional difference could be attributed to the different dust opacities if the dust size distribution is changed across a spiral due to e.g. shock waves.

\begin{figure}[ht]
    \centering
    \includegraphics[width=0.6\textwidth]{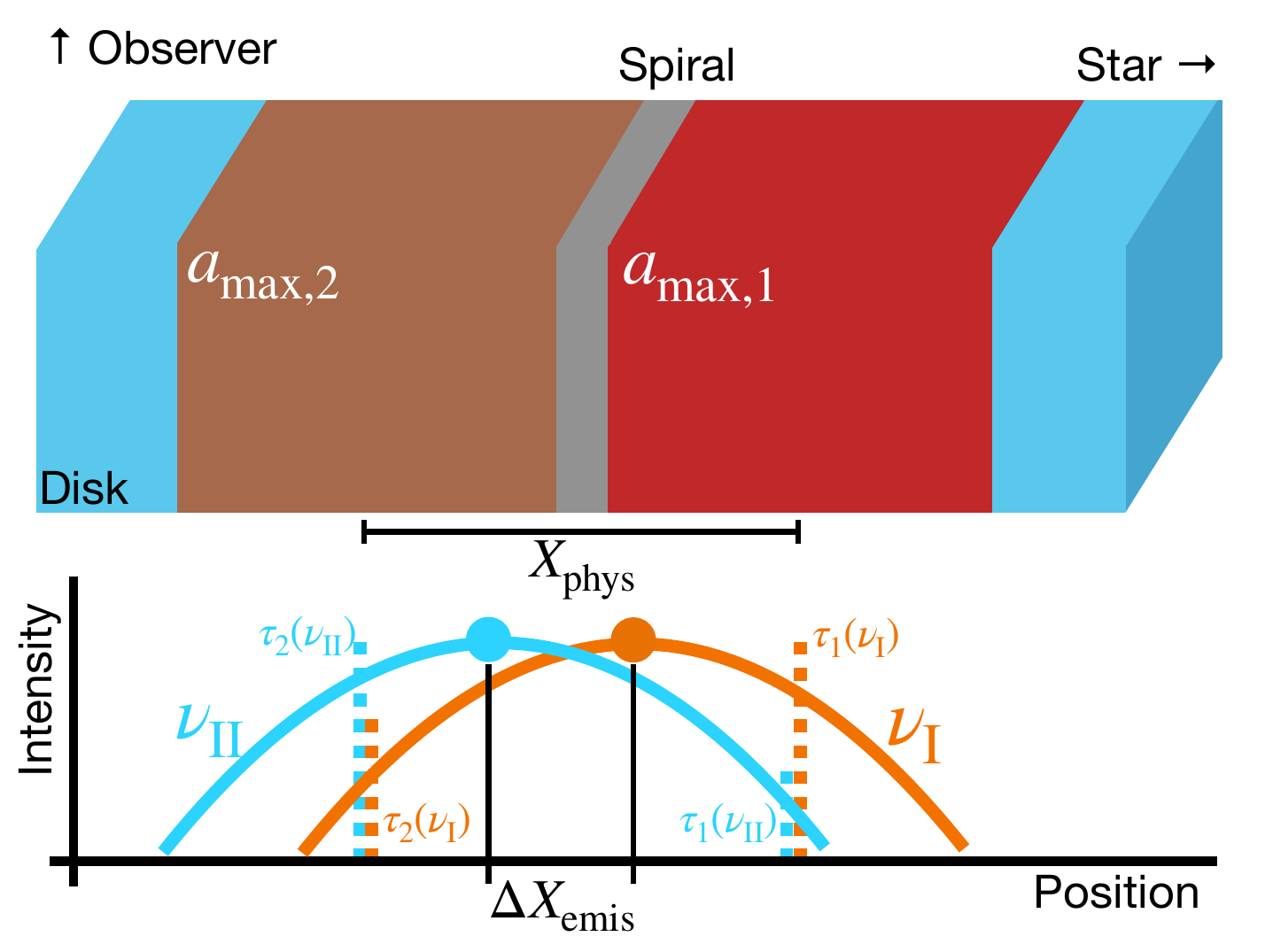}
    \caption{Schematic picture of the situation we consider. The dust size could differ across the spiral, which may cause the different opacity between the two frequencies. This could result in the apparent positional difference between the two frequencies even without the temporal variation. See the text for definitions of the parameters.}
    \label{fig:dust_il}
\end{figure}
To investigate this possibility, we consider a model with two homogeneous dust slabs behind and beyond a spiral, where dust size distributions are different as illustrated in Supplementary Figure \ref{fig:dust_il}.
Assuming that the dust size distribution is a power law with a cut off at the maximum dust size $a_{{\rm max}, i}$, where $i=1,2$ is the number of the slab behind and beyond the spiral, the optical depth from each dust slab can be described as
\begin{equation}
    \tau_i(\nu_j) = \kappa(a_{{\rm max}, i}, {\nu_j}) \Sigma_i,
\end{equation}
where $\kappa(a_{{\rm max}, i}, {\nu_j})$ is the dust opacity at the frequency $\nu_j$ $(j=\greone, \gretwo)$ and the maximum dust size $a_{{\rm max}, i}$, and $\Sigma_i$ is the dust surface density of slab $i$.
Under the condition where the two dust slabs are spatially unresolved, which is the case for this study, the observed position of the emission from the two slabs is their centroid.
If the two slabs are optically thin and have the same temperature, the centroid of the emission with respect to the position of the slab $i$ is given as
\begin{equation}
    X_{\rm emis}(\nu_j) =  \frac{ \tau_2(\nu_j)}{ \tau_1(\nu_j) + \tau_2(\nu_j) } X_{\rm phys},
\end{equation}
where $X_{\rm phys}$ is the physical seperation of the two slabs.
The centroid difference between the two frequencies can be expressed as
\begin{equation} \label{eq:dXemis}
    \Delta X_{\rm emis} = X_{\rm emis}(\nu_{\rm \gretwo}) - X_{\rm emis}(\nu_\greone) = \left\{ 
        \frac{ \tau_2(\nu_\gretwo)}{ \tau_1(\nu_\gretwo) + \tau_2(\nu_\gretwo) } - \frac{ \tau_2(\nu_\greone)}{ \tau_1(\nu_\greone) + \tau_2(\nu_\greone) }    
    \right\} X_{\rm phys}.
\end{equation}
The ratio $\Delta X_{\rm emis}/X_{\rm phys}$ suggests how the centroid of the emission changes between the two frequencies with respect to the physical separation of the two slabs.
Therefore, if this ratio can be as large as unity, the detected positional difference may also be interpreted as an opacity effect. On the other hand, if it is sufficiently small, the detected positional difference cannot be explained by the opacity effect but by the temporal variation.
For any combination of the dust sizes, $\Delta X_{\rm emis}$ is maximized when $\Sigma_1 = \Sigma_2$, so we can further write the upper limit of the ratio as
\begin{equation}
   \frac{\Delta X_{\rm emis}}{X_{\rm phys}} = \frac{ \kappa(a_{{\rm max}, 1}, {\nu_\greone})  \kappa(a_{{\rm max}, 2}, {\nu_\gretwo}) - \kappa(a_{{\rm max}, 1}, {\nu_\gretwo})  \kappa(a_{{\rm max}, 2}, {\nu_\greone}) }{ \{ \kappa(a_{{\rm max}, 1}, {\nu_\gretwo} ) + \kappa(a_{{\rm max}, 2}, {\nu_\gretwo}) \} \{ \kappa(a_{{\rm max}, 1}, {\nu_\greone} )- \kappa(a_{{\rm max}, 2}, {\nu_\greone}) \} },
\end{equation}
for a combination of the dust sizes and frequencies.

\begin{figure}[ht]
    \centering
    \includegraphics[width=0.8\textwidth]{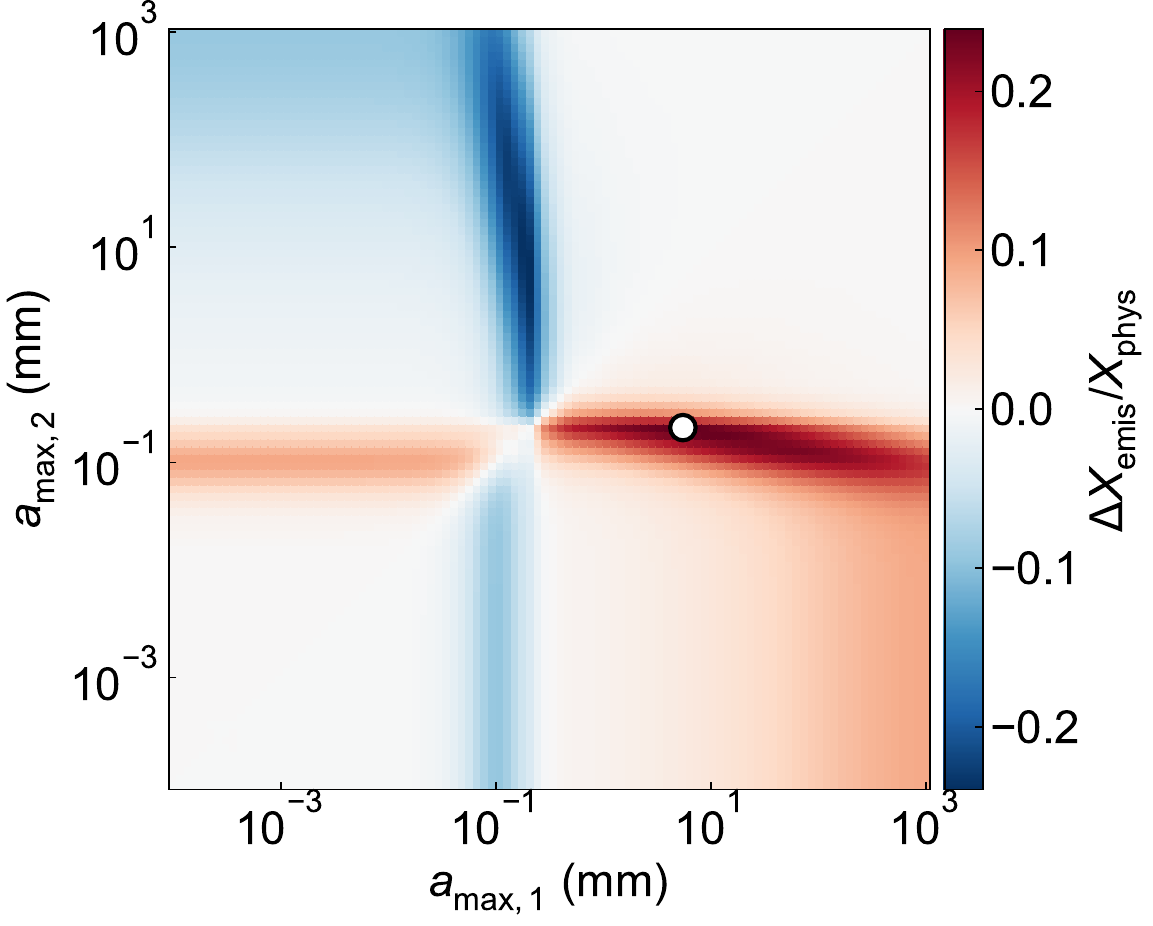}
    \caption{Ratio of the centroid difference of the emission between the two frequencies to the physical separation of the two slabs as a function of the maximum dust size of the two slabs. The white point indicates the maximum value (${\sim}0.24$), where $(a_{\rm max, 1}, a_{\rm max, 2}) = ( 5.5, 0.21 )\ {\rm mm}$.}
    \label{fig:dust-size}
\end{figure}
We calculated the ratio for the DSHARP dust opacity model \citep{birn2018} for various combinations of maximum dust sizes with $\nu_\greone = 240$ GHz and $\nu_\gretwo = 330$ GHz, which are the representative frequencies of 2017 and 2024 images (the longest time base-line), and plotted the results in Supplementary Figure \ref{fig:dust-size}.
The ratio is smaller than $0.1$ for a broad range of dust sizes.
For a limited parameter space, the ratio takes a somewhat large maximum of ${\sim}0.24$ when the maximum dust size of the two slabs are $(a_{\rm max, 1}, a_{\rm max, 2}) = ( 5.5, 0.21 )\ {\rm mm}$.
This means that, even if there is such a significant difference in the maximum dust size between the two slabs (i.e. across a spiral), the opacity difference can only make a positional difference of $0.24$ times their physical separation.

Based on our analysis, the apparent positional difference is ${\sim} 0\farcs05$ at 30 au from the central star.
If this were the frequency effect, the physical separation of the two dust components should be as large as ${\sim}0\farcs2$.
However, with our spatial resolution of $0\farcs15$, such a large physical separation should be spatially resolved, which is not the case (Figure 1 in the Main text).
In other words, the maximum physical separation that is possible for our spatial resolution only allows the frequency effect to be sufficiently smaller than actually observed. 
In addition, in the case of the frequency effect, it is required to satisfy very limited conditions, where the maximum dust sizes are fine-tuned and differ by a factor of ${\sim} 26$ with a comparable dust surface density.
Therefore, we conclude that the apparent positional difference is not due to the frequency effect but to the temporal variation.

\subsection{Supplementary figures and tables}
We summarize the parameters used for imaging (Methods \ref{obs}) in Supplementary Table \ref{tab:image}. 
The image-fitting (Methods \ref{image_fitting}) results are listed in Supplementary Table \ref{tab2}.
The marginal probability distribution is presented in Supplementary Figure \ref{fig:corner}.

\begin{table}[ht]
\caption{Imaging Details}\label{tab:image}%
\begin{tabular}{ccccc}
\toprule
Epoch & Weighting & {\it uv}-taper & Original beam & Sensitivity before convolution \\
& & & & ($\rm \mu Jy\ {\rm beam^{-1}}$) \\
\midrule
2017 & Super-uniform,  ${\rm \tt npixels} = 10$ & $0\farcs08$ & $0\farcs081 \times 0\farcs081,\ 41^\circ$ & 18 \\
2019a & Briggs, ${\rm \tt robust} =0$ & N/A & $0\farcs11 \times 0\farcs08,\ -81^\circ$ & 28 \\
2019b & Briggs, ${\rm \tt robust} =0$ & N/A & $0\farcs09 \times 0\farcs08,\ -88^\circ$ & 27 \\
2024 & Briggs, ${\rm \tt robust} =0$ & N/A & $0\farcs11 \times 0\farcs09,\ -77^\circ$ & 29 \\
\botrule
\end{tabular}
\end{table}

\begin{table}[ht]
    \caption{Parameters and fitting results}\label{tab2}%
    \begin{tabular}{cccccccc}
    \toprule
    Parameter & Prior & Percentiles & Best-fit & GI & Planet at 117 au & Unit \\
    \midrule
    $\log_{10} \omega_0$ & $\mathcal{U}(-4.7, 5.3)$ &  $-0.97^{+0.32}_{-0.39}$ & $-0.82$ & $-0.72$ & $-\infty$ & ${\rm deg\ yr^{-1}}$ \\
    $\log_{10} \omega_f$ & $\mathcal{U}(-4.7, 5.3)$ & $<0.04$ & $-3.19$ & $-\infty$ & $-0.53$ & ${\rm deg\ yr^{-1}}$ \\
    $\gamma$ & $\mathcal{U}(-5, 5)$ & $ 1.9^{+0.49}_{-0.44} $ & $1.75$ & $1.5$ & $0$ &  \\
    \botrule
    \end{tabular}
    \footnotetext{There are six other parameters on the parallel shift from the 2017 data. Their prior is set to $\mathcal{U}(-100, 100)$ mas. Since the marginal distributions are localized around zero and do not have any physical meaning, we omit them in this table. }
\end{table}

\begin{figure}[ht]
    \centering
    \includegraphics[width=\textwidth]{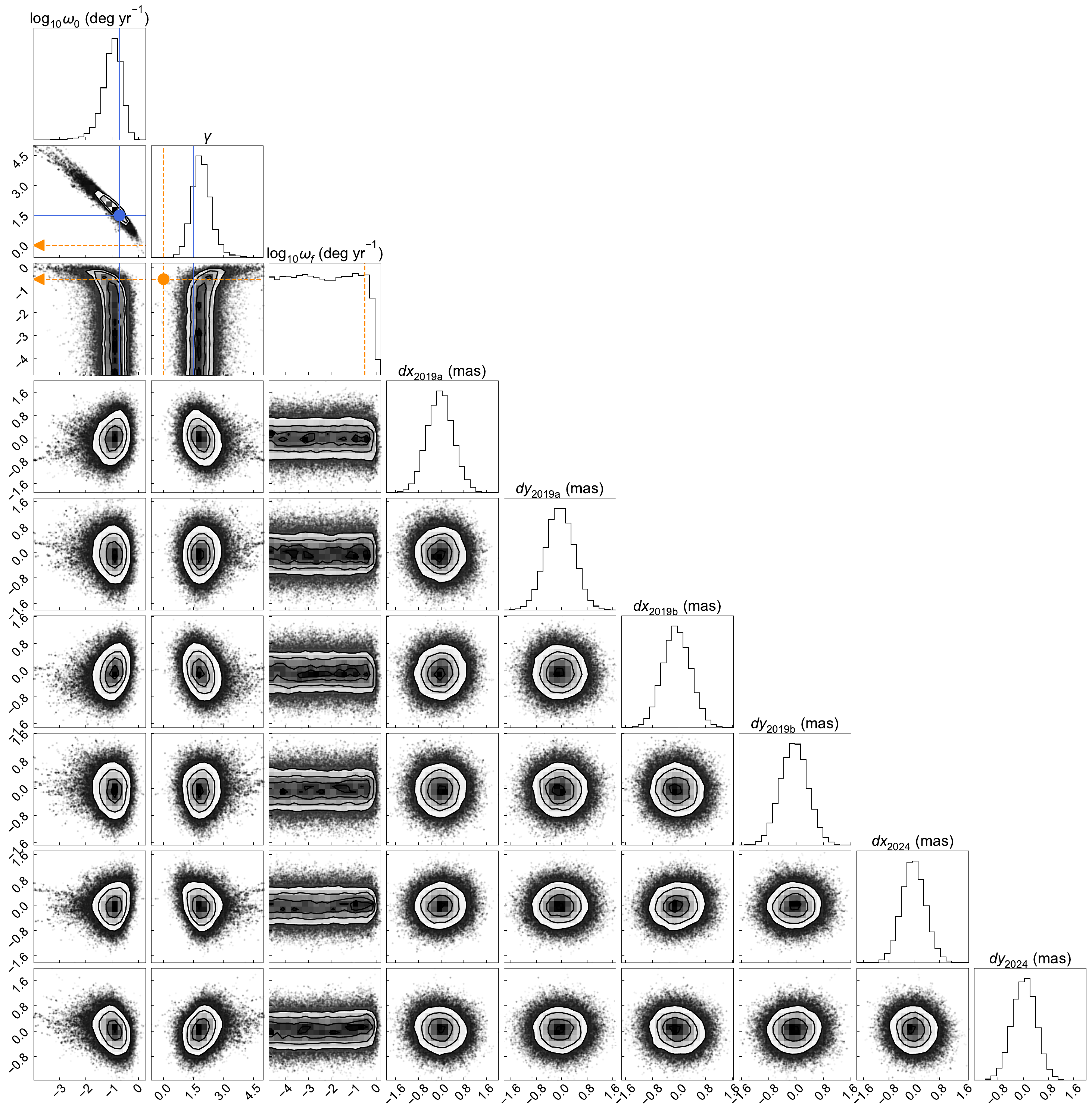}
    \caption{Marginal probability distribution. The blue and orange lines indicate the GI- and planet-induced cases, respectively.
    The GI case is consistent with the distribution, while the planet case falls in a region with almost no samples.
    }\label{fig:corner}
\end{figure}

\end{document}